\pdfoutput=1
\documentclass[annualconference]{acmsiggraph}  

\usepackage[scaled=.92]{helvet}
\usepackage{times}

\usepackage{graphicx}
\usepackage{subfigure}

\usepackage{parskip}

\usepackage[labelfont=bf,textfont=it]{caption}

\onlineid{0264}

\title{View-Dependent Displays and the Space of Light Fields}

\author{Roarke Horstmeyer\thanks{e-mail: roarkeh@mit.edu}\\ MIT Media Lab%
\and Se Baek Oh\thanks{e-mai: lsboh@mit.edu}\\ MIT Department of Mechanical Engineering%
\and Ramesh Raskar\thanks{e-mail: raskar@media.mit.edu}\\ MIT Media Lab}

\begin{document}

\teaser{
  \includegraphics[width=6.5in]{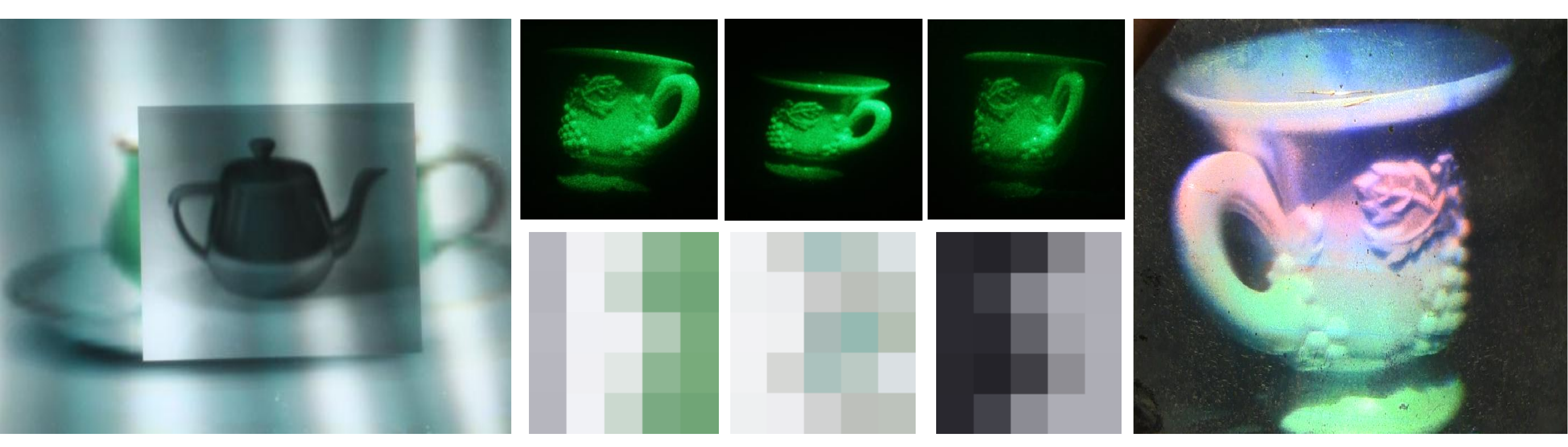}
  \caption{Parallax barriers and holograms are two forms of view-dependent display that are typically considered under distinct physical interpretations, but actually consist of similar underlying features. (left) A parallax barrier image comprised of different discrete tiles of angular content (center-bottom) are usually explained by treating light as a ray. By illuminating one discrete patch of a hologram at a time and recording the resulting images (center-top), we describe a hologram in light field terminologies commonly applied to parallax displays. In this paper, we develop a framework to explain these similarities and then apply it to the creation of novel display forms. }
}

\maketitle


\begin{abstract}

In this paper we explore how light propagates from thin elements into a volume for viewing. In particular, devices that are typically connected with geometric optics, like parallax barriers, differ in treatment with those that obey physical optics, like holograms. However, the two concepts are often used to achieve the same effect of capturing or displaying a combination of spatial and angular information. This paper attempts to connect the two approaches under a general framework based in ray space, from which insights into applications and limitations of both parallax-based and holography-based systems can be observed. We show that each display form can generate a light distribution that can always be expressed as a rank-1 matrix. Knowledge of this limitation is then discussed in the context of extending the capabilities of current display forms by considering the use of partially coherent light.


\end{abstract}


\section{Introduction}



There are numerous questions surrounding the operation of holograms from a computer graphics and display perspective that do not have an obvious answer. For example, can a hologram be represented as a light field? Can a hologram create any light distribution we want? Can we create a parallax barrier by taking strips of a hologram? What does it mean to say that a hologram creates a ``wavefront'', but a parallax barrier does not? Based on recent connections made between geometric and physical optics interpretation of light propagation by \cite{Zhang:09}, this paper suggests intial answers to these questions.

Parallax barrier displays were invented a little over a century ago, and holograms have been around for about half of that time. While both are used for the same purpose of presenting a combination of spatial and angular information upon illumination, there has been little comparison of their capabilities and tradeoffs. This paper brings the two forms of display under a common framework, and will explore what is possible at their intersection.

\begin{figure}[b!]
  \centering
  \includegraphics[width=1\linewidth]{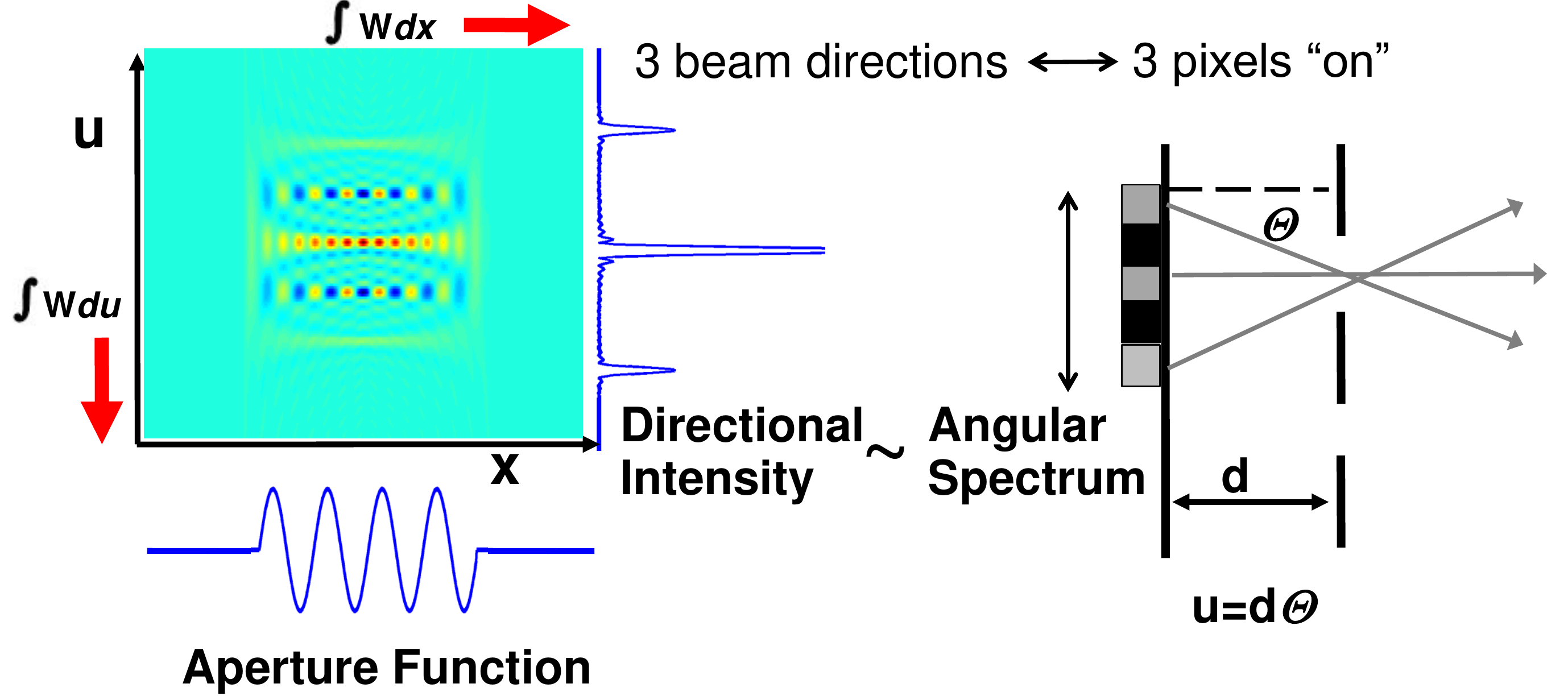}
  %
  \caption{
           The ray space diagram of a hologram (left) presents both spatial and directional distributions of the rays it emits. We show how this is equivalent to the angular spectrum of pixels underneath a parallax barrier slit (right), just at much smaller scales.}
\label{Fig2}
\end{figure}

\begin{figure*}[tb]
 \centering
  \subfigure{
  \includegraphics[scale=.35]{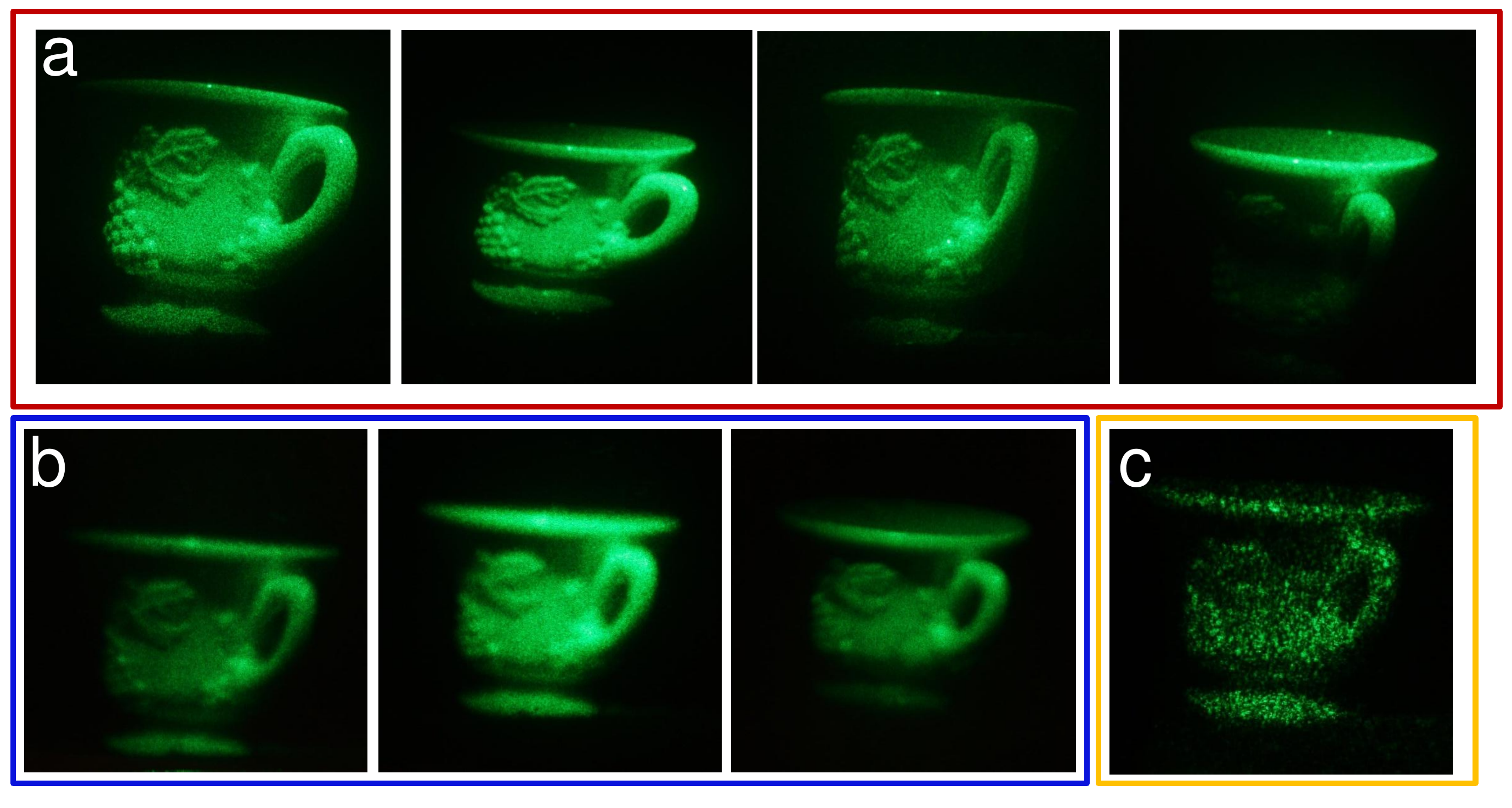}
	}
	\centering
	\subfigure{
  \includegraphics[scale=.25]{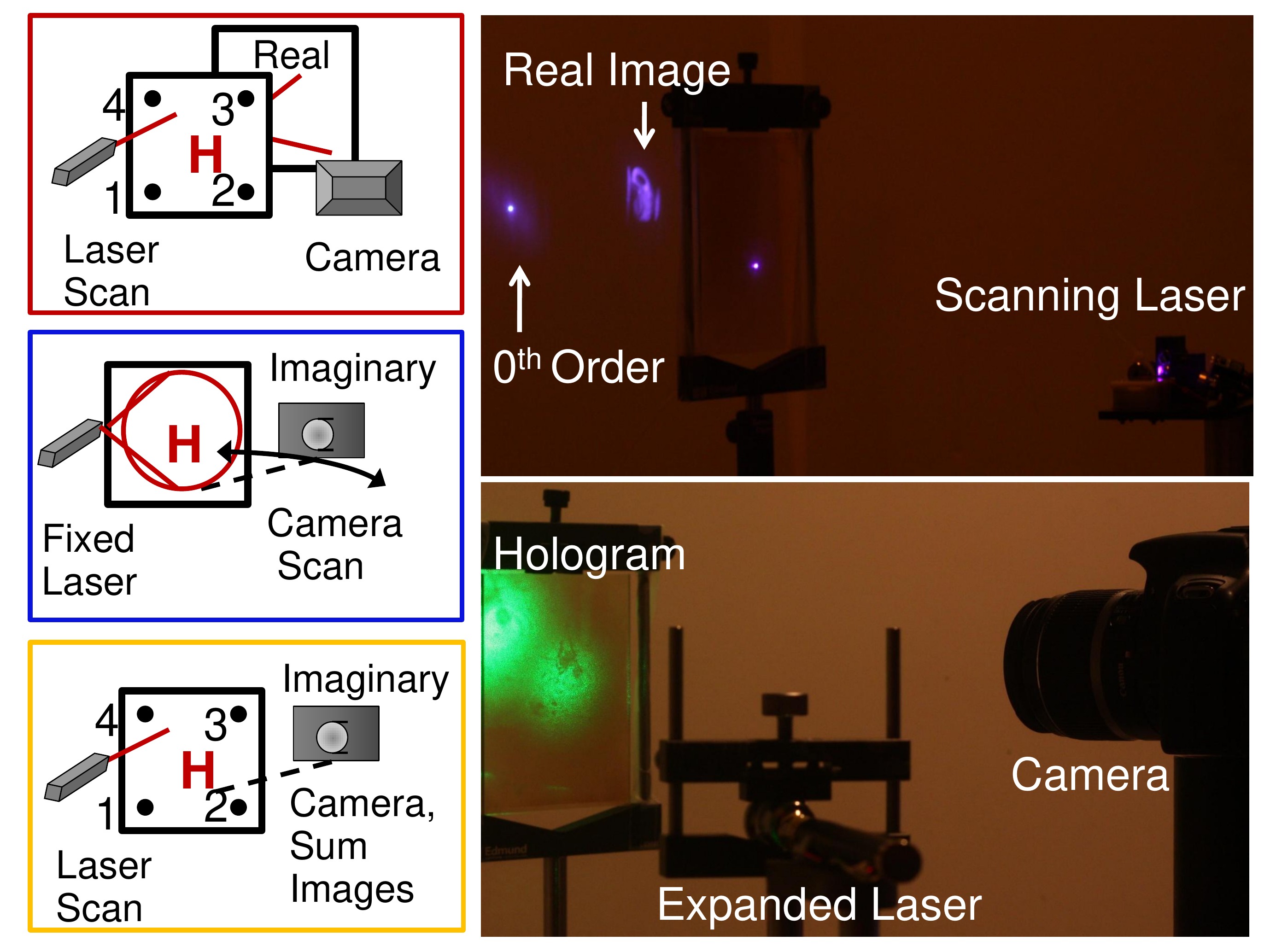}
	}
  \caption{\label{fig:firstExample}
            Experimental results for scanning and recording the 4D light field function of a hologram. (Left) Three unique recording methods each populate the light field. (a) The hologram is segmented into patches and scanned patch-wise, recording the real image. (b) A coherent source illuminates the entire hologram and the virtual image is captured from different angular positions. (c) The hologram is illuminated patch-wise and the virtual image is reconstructed through image addition. (Middle) The associated recording geometries. (Right) Images of the experimental setup for real image scanning and virtual image capture.}
\label{Fig8}
\end{figure*}

Parallax barriers, which are conceptually similar to lenticular arrays, are used to deliver the appearance of a three-dimensional image from a screen to a user without the need of special glasses. This screen is comprised of a plane of pixels with mixed spatial and angular content and a plane of slits. Holograms also contain a mixture of spatial and angular content. A traditional off-axis hologram preserves this mixture in the form of interference fringes, usually created from exposure to a plane wave and a second wave from an object of interest. Upon relighting with a plane wave, diffraction effects will create a real and virtual image of the object. If the object has an interesting 3D structure, this too will reappear in both the real and imaginary viewing volume.

There has been much exploration into the different methods of generating wavelength-scale amplitude or phase fringes that will diffract light to a desired pattern. In this paper, we will be mostly concerned with holograms that can be generated computationally. Furthermore, treatment will focus on holograms that can be presented in grayscale a thin transparent sheet. This specific holographic form is physically identical to the masks that make up a parallax barrier display.

It is well known that both holographic and parallax barrier displays are limited in the light distributions they can create. Using a linear algebra-based analysis, we will show that each form of display can only create light fields that are rank-1 in a certain space. Many people have attempted to solve the problem of coming up with the best parallax barrier or hologram design to approximate a desired light field. What they haven't realized is that they are attempting to satisfy the same rank-1 condition. By discussing the two display forms side-by-side and exploring their shared fundamental limits, we hope to encourage new areas of study at their overlap. Whats more, this overlap will only grow as pixel sizes shrink and the line between geometric and physical optics becomes less definite.

%
%
%
%
%

\subsection{Contributions}

As far as we know, this is the first time holograms and parallax barriers have been brought together under a joint geometrical and physical optics framework for direct comparison and analysis.We will present the following ideas in this paper:
\begin{enumerate}

\item  A side-by-side performance analysis of parallax barrier and holographic displays based upon discretizing spatial and angluar resolution.

\item A framework based in ray space to help show the limited space of light distributions that can be created by any thin amplitude element in a volume. We will show that this space is always rank-1 in the transform domain. 

\item The experimental capture of the light field of a hologram, demonstrating that a hologram can simply be thought of as a discrete summation of elemental patches that present individual viewpoints of a certain object. 

\end{enumerate}

\begin{figure}[b!]
  \centering
  \includegraphics[width=.95\linewidth]{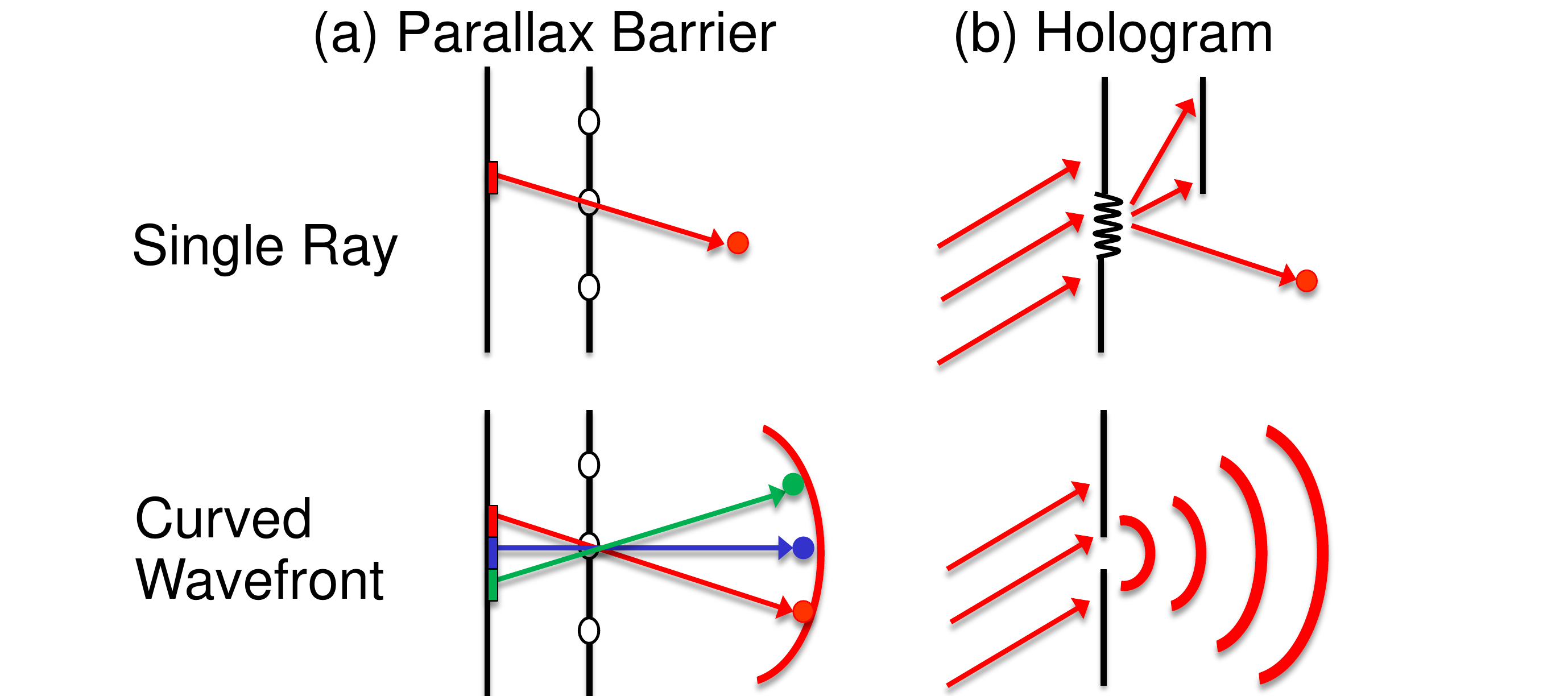}
  \caption{
           (Left) A parallax barrier can easily create a specific ray from a single pixel and a slit, but can only create a discrete approximation to a curved wavefront. (Right) An amplitude hologram can create the same ray with a sinusoidal pattern, but will generate two additional orders. It can easily create a curved wavefront with a narrow slit.}
\label{Fig4}
\end{figure}

\section{Related Work}

The term ``hologram'' has been applied to a wide range of media over the past sixty years, and this paper does not intend to consider all of them. The first genre of holograms was a film-based media~\cite{Gabor:48,Leith:65}, having different properties if thick or thin, or phase or amplitude~\cite{Urbach:69}. In this paper, we will be primarily concerned with holographic patterns that can be generated on a computer, some of the first of which were presented by Lohmann~\shortcite{Lohmann:67}. Since these early efforts, computer-generated holograms (CGH) have opened up a large field of study in their application to 3D  display (\cite{Benton:06}, \cite{Slinger:05}). However, many of the current advanced techniques, often based on segmenting holograms into zones, have roots in earlier film-based work~\cite{DeBitetto:69,Benton:69}). 
 
CGH's have been extended to model intensity distributions at multiple planes~\cite{Dorsch:94} and into continuous volumes~\cite{Piestun:96,Shamir:02}. Spatially--multiplexed holograms, like a holographic stereogram~\cite{McCrickerd:68}, display many discrete viewpoints of an object similar to a parallax barrier display. They too often exhibit only parallax along the horizontal direction, but not always, and provide a user with the appearance of a fully 3D object~\cite{Lucente:94}, like other advanced holographic displays ~\cite{Hilaire:91}. A helpful introduction to the different methods of encoding a desired display field onto different two dimensional light modulators can be found in~\cite{Ples05}. 
  
The first parallax barrier setup, developed by Ives~\shortcite{Ives:03}, was a binocular display delivering two slightly different image perspectives to each eye. 
A simple alternative to a parallax barrier is a lenticular display, first suggested by Lippmann~\shortcite{Lippmann:08}. In-depth comparisons between lenticular, barrier and integral systems can be found in~\cite{Okoshi:76} and~\cite{Javidi:09}, and a physical optics perspective is in ~\cite{Moller:05}. We will treat all of these two-plane systems in a simplified manner, but under a ray--space based framework similar to Zwicker et. al \shortcite{Zwicker:06}, where parallels to holography are close.  
  
Considerable work has also been focused on merging wave-based holography with geometrical frameworks~\cite{Zhang:09,Ziegler:07,Oh:09} Several works have drawn some simple comparisons between what is possible with geometric-based and holographic displays~\cite{Halle:97,Frauel:04} even applied to graphics applications ~\cite{Escriva:07}. Others have integrated display forms from the two genres~\cite{Bimber:04,Plesniak:03}. We hope to expand on these general comparisons using ray space analysis like the light field~\cite{Levoy:96}. The Wigner distribution~\cite{Wigner:32} is another ray space function that has many applications in physical optics~\cite{Bastiaans:79}. Holograms in specific have been analyzed from both of these ray space perspectives in~\cite{Testorf:08} and~\cite{Situ:07}, but not directly compared to parallax barriers in the same space.


\section{Towards a Common Framework}
 
Two key differences between the display media should be kept in mind while our comparison is formulated. First, parallax barriers operate in the geometric optics limit, while holograms require the physical optics principle of diffraction. This fundamental difference is related to the scale of the display pixel. Second, a simple parallax barrier display directs unique images to different angles, essentially mapping $\theta$ to $x$ at a single depth plane. The operation of a hologram, on the other hand, is a little bit more complex, since it can create both a real and virtual image. We will see that a hologram's virtual image is conceptually equivalent to that presented by a parallax barrier, while its real image acts as a resolution conjugate.


\begin{figure}[t!]
  \centering
  \includegraphics[width=.9\linewidth]{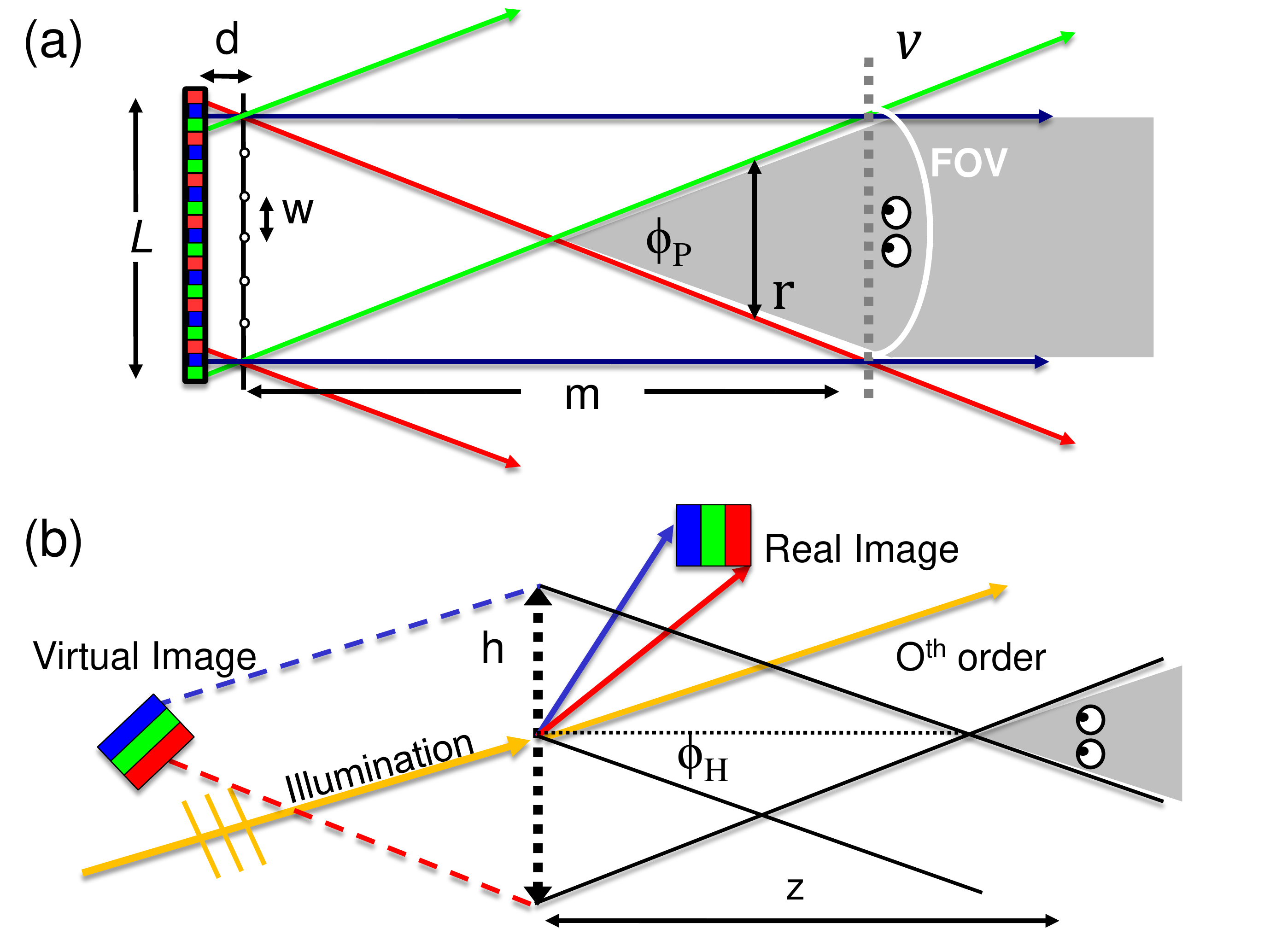}
  \caption{\label{fig:firstExample}
           (a) The operation of a parallax barrier display. A screen of width $L$ and a second plane of slits separated by $w$ a distance $d$ away directs light into specific directions. (b) Coherent illumination of a hologram creates both a real and virtual image through diffraction from a single plane. A viewer at a distance $z$ will see the virtual image appearing on the other side of the hologram. }
\label{Fig5}
\end{figure}

\subsection{Parallax Barriers Versus Holograms} 

The most generic parallax barrier configuration contains two planes with no additional optical elements, and light from each pixel only travels through one slit (Figure 5(a)). 
We focus on parallax barriers, and not lenticular or lenslet arrays, for two reasons. First, having both displays simplify down to thin amplitude modulation sheets underscores similarities. Second, as parallax barriers become increasingly based on LCD technology, which are becoming well above hundreds of DPI, the two display forms are converging in resolution. Table 1 summarizes many of the main properties of this simple parallax barrier along with holograms.

\begin{table}[t!]
  \centering
  \includegraphics[width=.8\linewidth]{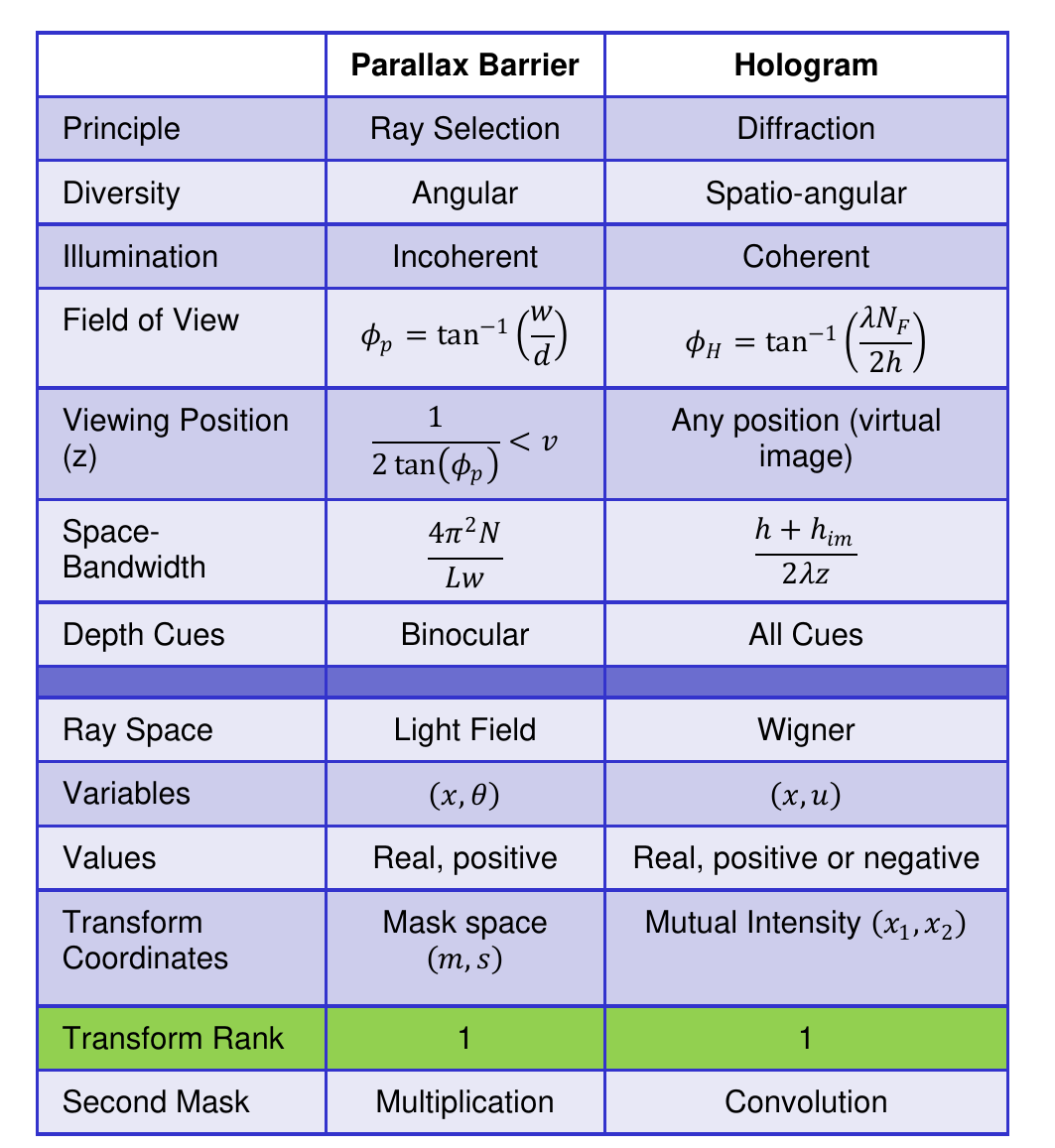}

  \caption{\label{fig:firstExample}
           Direct comparisons between parallax barrier and holographic 3D displays. The top section of the table outlines differences in implementation. The bottom section of the chart outlines the similarities and differences in the ray space treatment of each.}
\end{table}

For a parallax display with a given number of pixels $N$, it is clear that there is a direct tradeoff between the amount of spatial and angular content that can be directed to an optimal viewing plane $v$. Specifically, angular resolution ($\theta_{res}$) can be given by the number of pixels under each slit, and spatial resolution ($x_\mathit{res}$) as the number of slits, yielding 
\begin{equation}
x_{res}\theta_{res}=N.
\label{eqn1}
\end{equation}
At optimal viewing of this simplified display one ray from each slit will enter one eye, and a discrete number of of views are visible from different positions along $v$.

This discretization is one of the main drawbacks of a parallax barrier, and leads to issues like aliasing and orthoscopic views. High angular resolution is desirable to create a more seamless viewing experience. However, parallax barrier displays scale poorly with an increase in resolution for a fixed display size, mostly as a result of two issues. The first is a decrease in light efficiency, as optimal slit width is roughly the width of one display pixel. Thus, efficiency is decreased proportional to an increase in angular resolution for a fixed pixel size. A lenticular array can be used to improve this efficiency, but it will not completely overcome the second issue of diffraction. A slit of width $s$ will diffract a ray across an angle given by
\begin{equation}
\sin{\theta} =\frac{\lambda}{2s}
\label{eqn2}
\end{equation}
Likewise, a lenticular with f-number $F$ will be able to display a pixel with a minimum size given by the Airy disk as $x=1.22\lambda F$. Thus, as systems scale towards smaller pixel and barrier widths, physical optics effects cannot be ignored.

Diffraction is exact how a hologram achieves image creation. The thin transmission holograms we are considering here will operate under ``conventional'' holographic methods, in which light from all points in the holographic plane diffracts and interferes to produce each and every image point in a continuous manner. Many holograms do not obey this assumption, and by taking advantage of the fact that depth cues primarily come from horizontal parallax, can significantly reduce required resolution. The well-known rainbow hologram visible under white light ~\cite{Benton:69} and holographic stereograms ~\cite{McCrickerd:68} are examples that can enhance viewing conditions using techniques related to but beyond the scope of this paper.
 
How can we create a single ray or finite beam with a hologram? From Figure \ref{Fig4}, we see that in the simplest form, a single pixel and a slit can be replaced with a small sinusoidal grating and a barrier to block two of the three diffraction orders. The width of the ray this sinusoidal grating creates will not be finite, however, but is given by $\Delta\theta_{s}=\frac{\lambda z}{h_{0}}$, where $z$ is the image distance and $h_{0}$ is the grating width. A basic Fourier hologram, which creates a single real image from one perspective, can be thought of as a summation of sinusoidal gratings that diffract light into different viewing directions, much like a parallax barrier.  The image from a Fourier hologram is in the ``far-field'', where light propagates to a pattern that is a scaled Fourier transform of the mask design. 

A Fresnel hologram, unlike a Fourier hologram, produces the familiar real and virtual images that change depending upon perspective. These operate under the assumption that the object to be reconstructed is in the ``near zone (Fresnel zone)'', or relatively close to the holographic mask. Upon coherent plane wave illumination, a conventional hologram creates a real image and imaginary image at locations displayed in Figure 5(b). These images, unlike a parallax barrier, do not exhibit discrete angular content, and can provide full depth cues to a viewer. For the next two sections, we will focus on the Fresnel hologram's real image. The virtual image will be discussed in Section 3.4.


\begin{figure}[t]
  \centering
  \includegraphics[width=.95\linewidth]{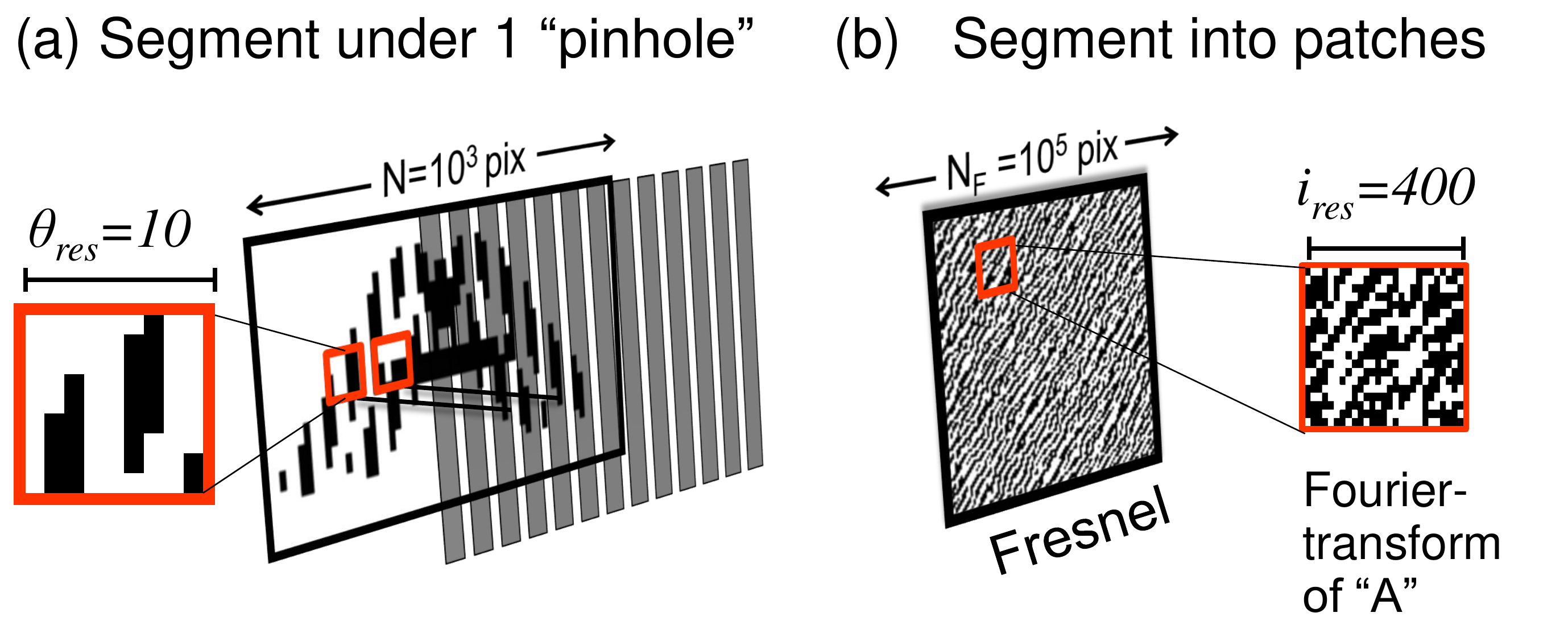}

  \caption{
           The discretization of a Fresnel hologram into a series of Fourier holograms allows it to be treated similar to a parallax barrier. Taking numerical examples from the successful display scenarios in Figure \ref{Fig6}, we can establish the resolution of a strip of a parallax barrier (a) will be $\theta_{res}=10$. For a Fresnel hologram (b) with $10^{5}$ pixels, the associated $i_{res}$ will be 400. 
}
\label{Fig5}
\end{figure}

\subsection{Discretization of the Hologram}

Although not exact, a convenient way to understand the operation of a Fresnel hologram is to split it into small, discrete patches, similar to what is done with a parallax barrier (Figure \ref{Fig5}).  Each holographic patch operates as an independent Fourier hologram, which tile together to create the complete 3D light distribution upon illumination. Thus, each of these elemental Fourier holograms projects its own slightly different perspective of the 3D object to create a real image. 

The number of required pixels to fully reconstruct the entire parallax content in a Fresnel (near-zone) hologram for an image of size $h_{im}$ is
\begin{equation}
N_{F}=\frac{h(h+h_{im})}{\lambda z}.
\label{eqn4}
\end{equation}
Likewise, a good rule of thumb for the generation of a Fourier hologram is that it must have 4 times the resolution of the real image, due to the creation of multiple orders (i.e., the sinusoid mask always creates more than 1 ray). Using these two resolution approximations, the number of elemental patches $n_{p}$ that a large Fresnel hologram will contain given a desired spatial resolution $i_{res}$ for the  reconstructed object is,
\begin{equation}
n_{p}=\frac{N_{F}}{4i_{res}}.
\end{equation}
 This equation is quite similar to (\ref{eqn1}) that describes a resolution tradeoff of a parallax barrier. It is very easy to draw a direct parallel between the number of patches $n_{p}$ and $\theta_{res}$ of the parallax barrier when considering a hologram's real image.

\begin{figure}[t]
  \centering
  \includegraphics[width=1\linewidth]{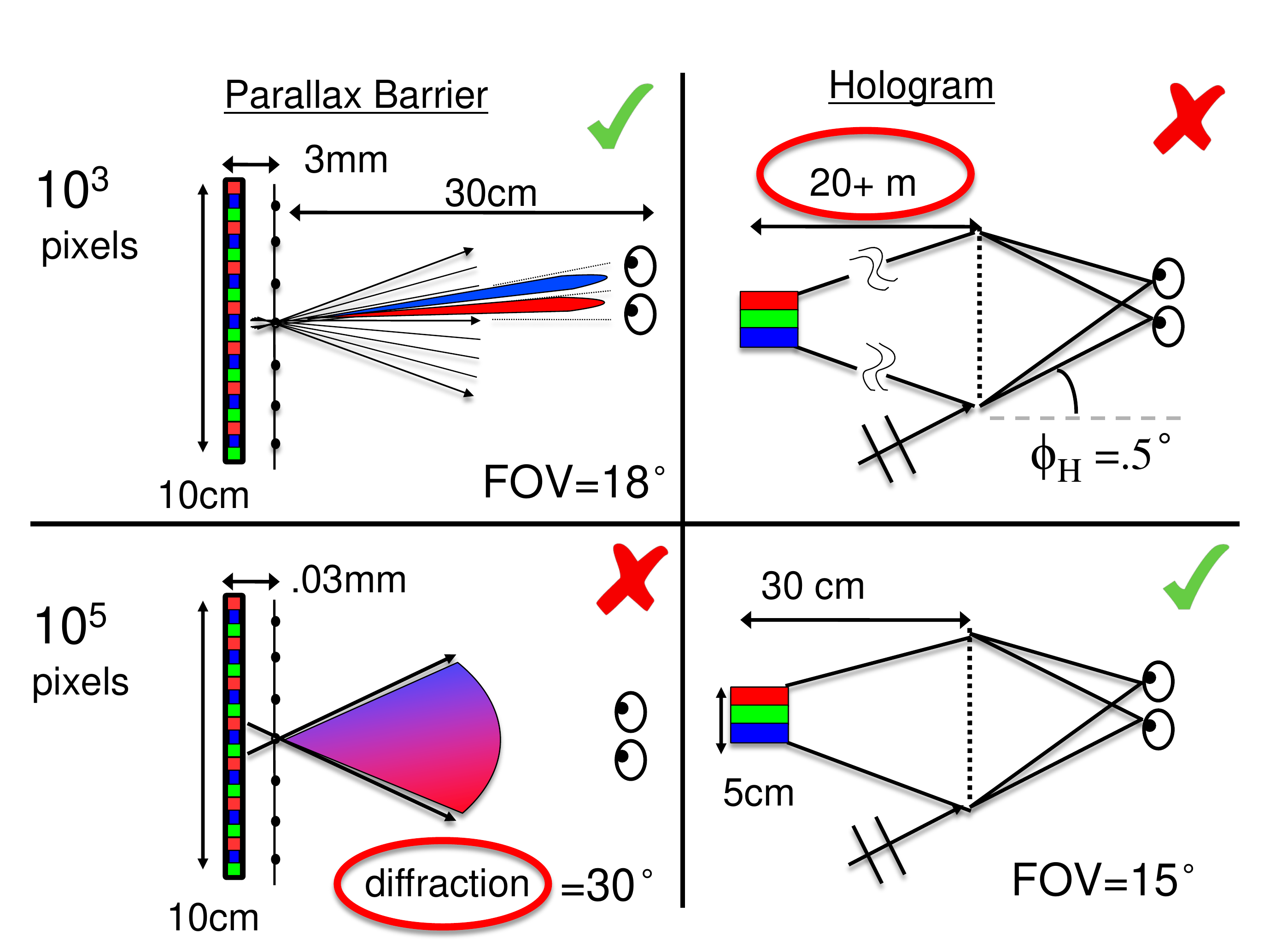}

  \caption{\label{fig:firstExample}
           A numerical comparison between parallax barrier and a holographic display shows that the barrier display performs well with larger pixels ($100~\mu$m), while a hologram performs well with smaller pixels ($1~\mu$m).}
\label{Fig6}
\end{figure}

\subsection{A Numerical Comparison}

A direct numerical comparison of the two display technologies helps clarify their areas of operation and the idea of treating a hologram in discrete patches. Figure \ref{Fig6} displays the operation of a parallax barrier and hologram for a screen with either $10^{3}$ or $10^{5}$ pixels fit into 10 cm. The parallax barrier is successful with $10^{3}$ pixels that are 100$\mu$m wide, consistant with current slit widths ~\cite{Harold:03}, but suffers from diffraction using smaller pixels. For the successful parallax barrier, the screen was split up such that $x_{res}=100$ and $\theta_{res}=10$. 

A hologram performs better with $10^{5}$ pixels that are 1 $\mu$m wide, becuase it can diffract light across a larger angle ($30^{\circ}$ compared to $0.5^{\circ}$, using (\ref{eqn2})). From (\ref{eqn4}), we find that this setup can fully reconstruct all of the parallax information contained by an object that appears 30 cm away and is 5 cm high.
  
Splitting up the hologram into discrete patches allows us to imagine how this successful CGH could be designed. To create a real image with 100 pixels of spatial resolution like successful barrier display, each Fourier hologram patch must have roughly 400 pixels (Figure \ref{Fig5}). The mask can be split up into 250 elemental components, yeilding 250 unique perspectives. This same process will next be applied to describe the resolution of the hologram's virtual image, which will link it to the image created by a parallax barrier.

\begin{figure}[t]
  \centering
  \includegraphics[width=.95\linewidth]{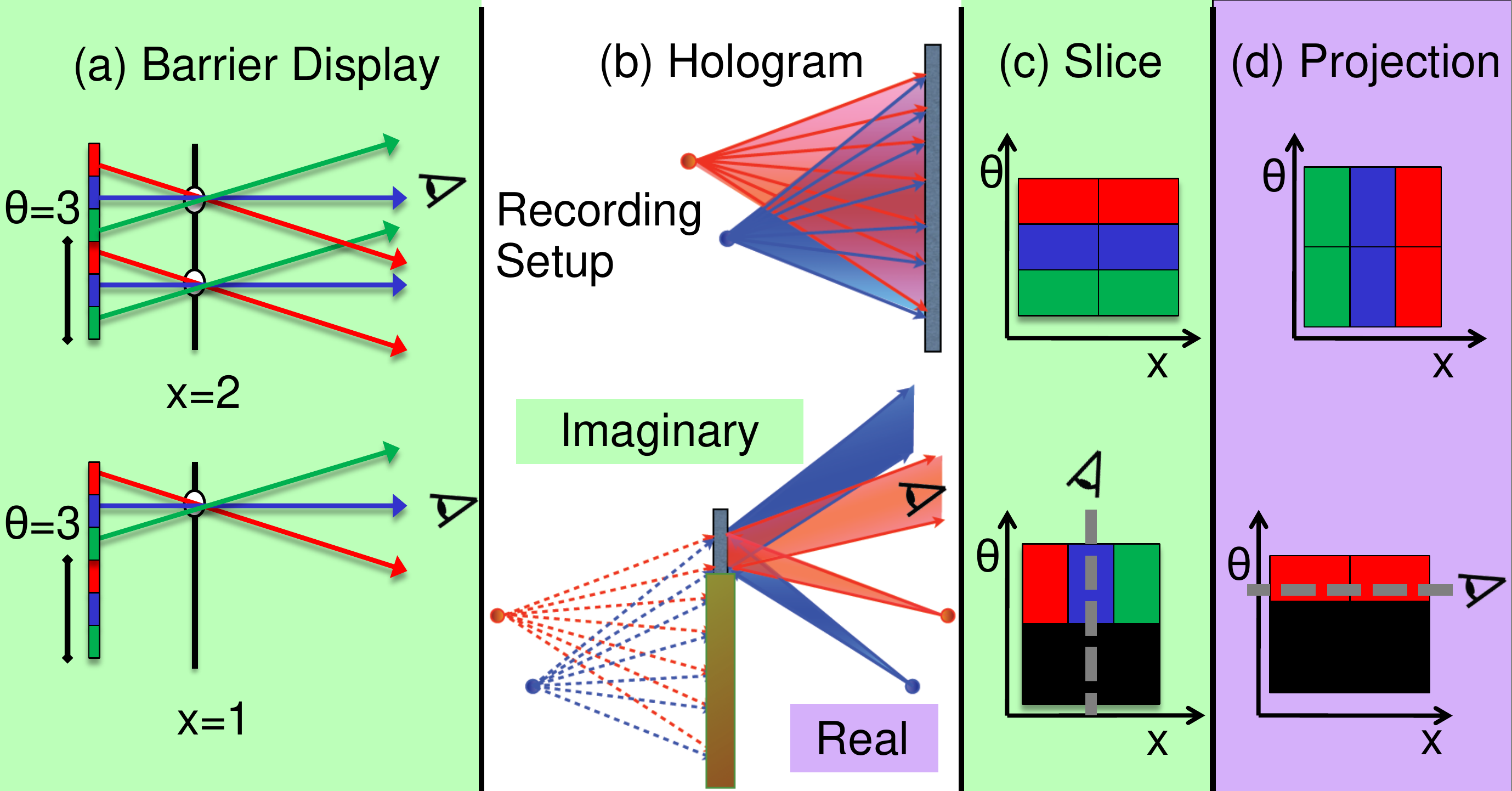}

  \caption{\label{fig:firstExample}
           The virtual image of a Fresnel hologram is conceptually the same as the image from a parallax barrier, and the real image is a resolution conjugate. Blocking off large regions of each display demonstrate this. (a) One pinhole of a parallax barrier will just give the angular content from one local region of the image. (b) Blocking most of a hologram of two points leads a partially obscured virtual image, giving angular content from only one point, just like in (a). The real image will still be fully visible. (c) A slice along the light field's $\theta$ can decribe the operation of a parallax barrier and imaginary hologram image. (d) A projection along $x$ describes a holographic real image.}
\label{Fig7}
\end{figure}

\subsection{Resolution and the Real and Virtual Image}

Holographic virtual images are viewed by looking ``through'' the hologram mask (Figure 5(b)). Thus, the 250 discrete Fourier patches that fit in our example in the previous section will now switch to define the spatial resolution of the virtual image, much like the slits of a parallax barrier. The 400-pixel resolution of each patch will also switch from providing the spatial resolution of the real image to the angular resolution of the virtual image. In summary, a hologram's spatio-angular tradeoff is connected to whether the real or virtual image is being viewed, as the two are resolution conjugates.

The light field point of view, a joint ($x, \theta$) representation of the displayed rays, is helpful at explaining this example. A real image is a projection of the light field along the $x$--axis, while a virtual image is a slice along the $\theta$-axis, as in (Figure \ref{Fig7}.c). Thus, either multiple real or virtual images may be recorded to define the light field of a hologram.

\subsection{Experimental Demonstration}

Using these new insights, the experimental capture of a hologram's light field is now accomplished using three different procedures. First, we capture unique real-image views by illuminating different locations of the hologram. This is identical to recording the low resolution image produced by each discrete Fourier patch ($x_{h}$,$y_{h}$) of a single angular view ($u_{im}$,$v_{im}$).  Scanning the illumination to multiple spots populates the entire light field, angle by angle. This hypothesis is verified by discretizing a recorded silver halide Fresnel hologram of a tea cup into 5 mm $\times$ 5 mm patches (Figure 3(a)). Each patch acts as a single hologram spatial coordinate, although they individually contain an estimated $1000^{2}$ 5 micron phase pixels. This resolution maps to roughly a 250 square pixel image resolution with the 4-1 resolution rule of thumb. An automated laser scanning system captures 400 images to reconstruct a (250,250,20,20) hologram light field.   

The virtual image is used to reconstruct the object two ways. In one method, the entire hologram is illuminated with a beam-expander and the camera moves to different angular positions for unique views (Figure 3(b)). Now similar to a parallax barrier display, the discrete hologram patch coordinates become the spatial resolution and camera position defines the angle ($u_{im}$,$v_{im}$).

In the final method, the hologram is illuminated in discrete patches and viewed as a virtual image, but with the camera initially fixed (Figure 3(c)). Each patch($x_{h}$,$y_{h}$) is now responsible for a single angular ray in the virtual image space. An entire holographic recording is formed by moving the illumination to different patches and summing images.    

Besides demonstrating the close link between holograms and light fields, this experiment serves as a method to digitally preserve traditionally recorded holograms, or even as a method to integrate lighting position into more complex displays. Qualitatively, scan-recording the real image produces the best results, although it is an unconventional technique and to the best of our knowledge has not been applied before.

\section{Space of Light Fields and Display Devices}

In this section, we will first show through a linear algebra analysis that the space of 3D light distributions from a parallax barrier and from a single holographic mask are both Rank-1. Then, the entire spectrum of achievable light fields (Figure \ref{Fig9}) ranging from coherent to incoherent light is explained through example to encourage new insights into the design space for 3D displays. We will then apply these insights in Section 5 towards the approximation of an inverse problem and the creation of a novel display format. To achieve this, it will be useful to transfer the problem to a joint configuration space comprised of spatial and spatial frequency coordinates. Following is a brief introduction to this representation and its comparison to light field space.

\begin{figure*}[htb]
 \centering
  \subfigure{
  \includegraphics[scale=.24]{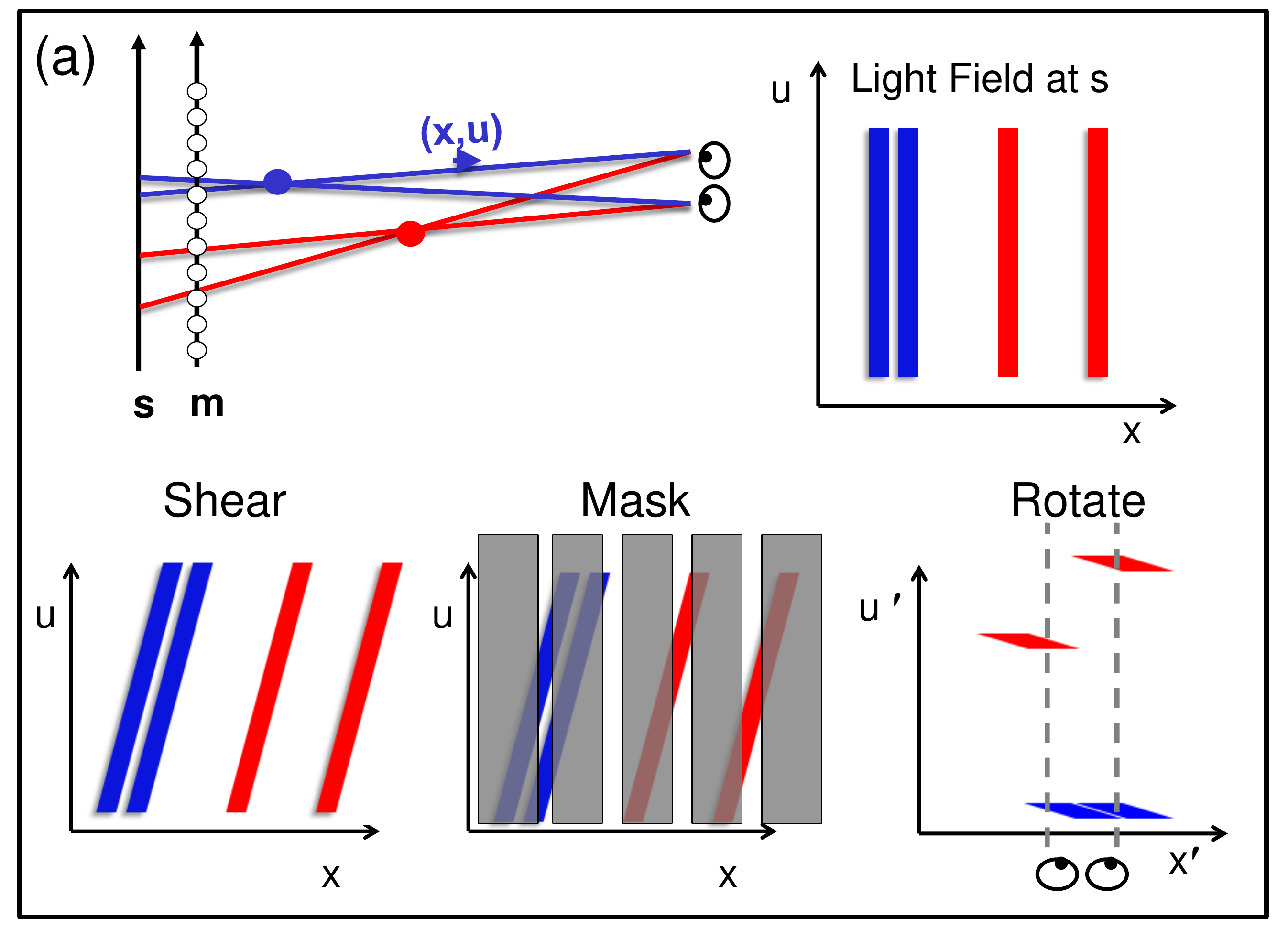}
	\label{Fig10a}
	}
	\subfigure{
  \includegraphics[scale=.24]{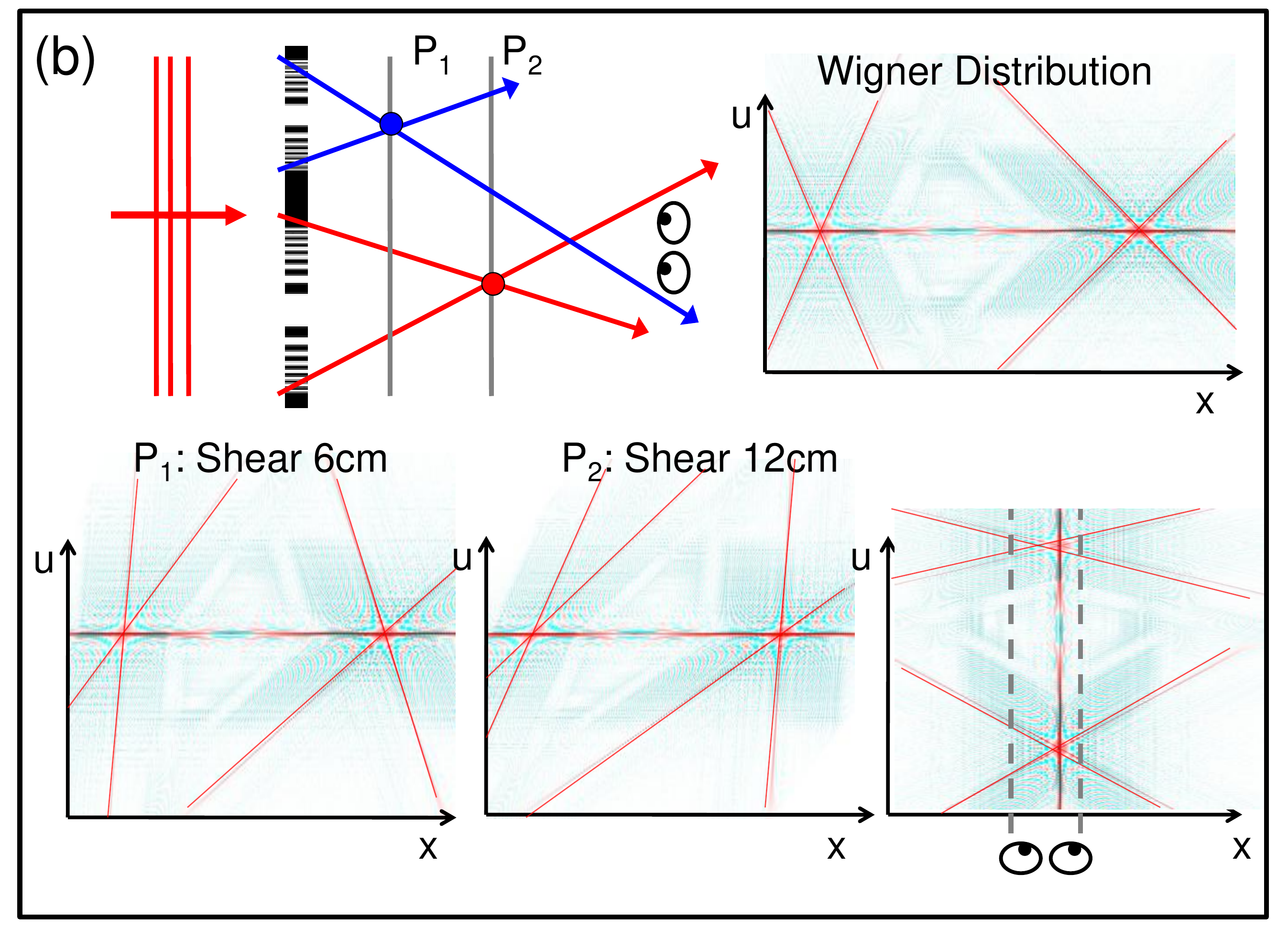}
	\label{Fig10b}
	}
	\subfigure{
  \includegraphics[scale=.25]{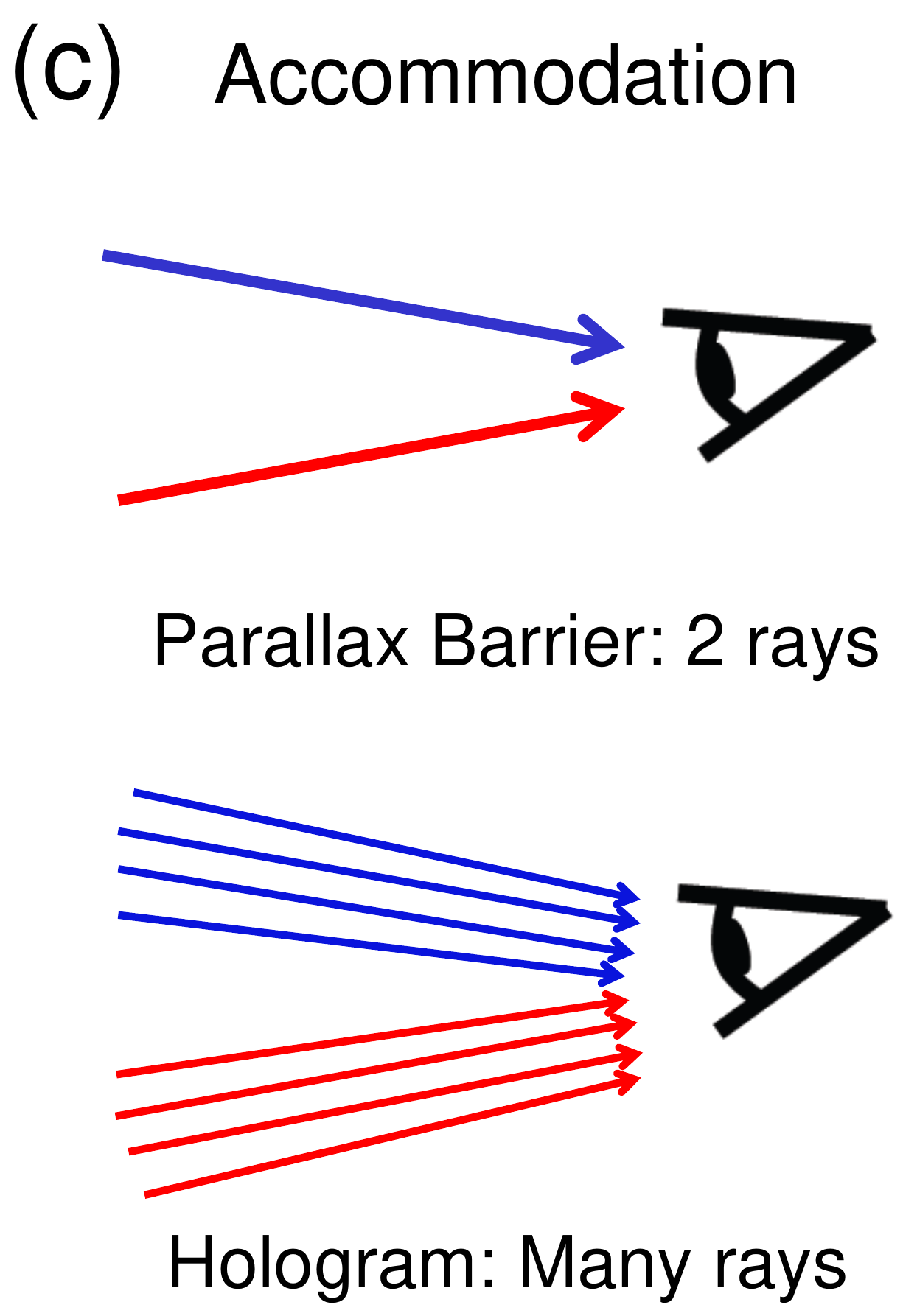}
	\label{Fig10c}
	}
  \caption{\label{fig:firstExample}
           Ray space representations for a parallax barrier (a) and holographic (b) display in an example where both attempt to create a three-dimensional image of two points. The parallax barrier will have 4 pixels create a broadband light field, which shears to a mask plane and is attenuated. A viewer sees the rotated version of this attenuated light field. A holographic system will use two Fresnel zone plates to create two points. The Wigner distribution from this mask will go through identical shears and rotations to a viewer. (c) The large number of rays entering a viewer's eyes with the hologram facilitates accommodation. }
\label{Fig10}
\end{figure*}


\begin{figure}[tb]
  \centering
  \includegraphics[width=.95\linewidth]{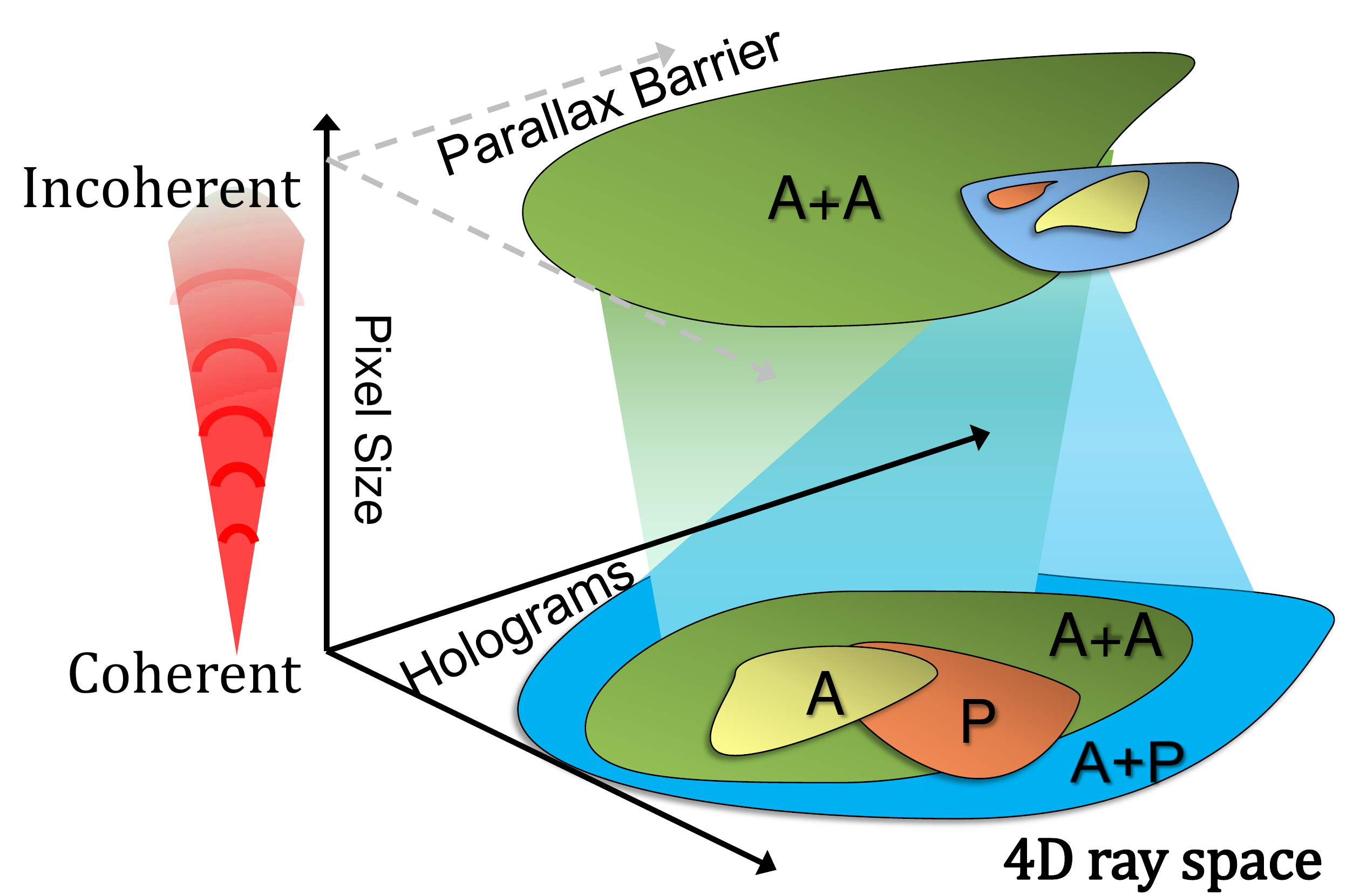}

  \caption{\label{fig:firstExample}
           The space of 4D light distributions. With coherent light, one mask with total control over amplitude and phase (A+P) can create any physically realizable wavefront. The size of this space is rank-1 in mutual intensity space. 
This is the type of mask we will examine in Section 5. With incoherent light, two amplitude masks, like a parallax barrier, can occupy the largest space, as the utility of phase becomes diminished. The size of this space is also rank-1. }
\label{Fig9}
\end{figure}

\subsection{Parameters of Ray Space}

The parameterization of light into a joint position-angle ray space $l(x,\theta)$ is well known. In this space, propagation is represented as a shear, ray values are always positive, and rays cannot bend upon interference with a mask. First considering a parallax barrier (Figure \ref{Fig10a}), the initial screen of pixels creates a broadband light field $l_{1}(x,\theta)$. This then shears and interacts with a plane of amplitude modulating slits, which we will call $l_{2}(x)$. The slits will either block or allow a ray through, leading to an expression for the final light field in terms of a multiplication:
\begin{equation}
l_{3}(x,\theta)=l_{1}(x,\theta) l_{2}(x).
\label{eqn6}
\end{equation}
This light field then shears across a large distance to a viewer's eye, which takes a projection along $\theta$.

To shift our analysis to masks with pixel sizes close to the wavelength of light, we must transform the light field into something that works with physical optics, or waves. Just as numerous rays can be added together to form a light field, so too can any wavefront be represented as a summation of plane waves of different spatial frequencies ~\cite{Goodman:96}. The Wigner distribution (Fig. 9(b)) is a function that simultaneously presents this spatial and local spatial frequency information of a monochromatic wavefront in a single plot. In geometric terms, this distribution can be simply thought of as a light field distribution with the ray angle $\theta$ replaced with the spatial frequency of its associated wavefront: $u=\frac{\sin\theta}{\lambda} \approx \frac{\theta}{\lambda}$. 

It is well known that in the limit of a very small wavelength, physical optics-based functions like the Wigner distribution approach radiance functions, or rays ~\cite{Accardi:09,Foley1985236}.  This implies that the light field and Wigner distribution are applicable at different scales, like the two displays we are discussing. A light field, like a parallax barrier, works well when variables are much larger than the light wavelength. If we zoom into the system and look at pixels close to the size of a wavelength, as in a hologram, we simply need to think of rays as plane waves. 

At small scales, if a transparency is illuminated by a coherent plane wave parallel to the z-axis, then the resulting ray space diagram after passing through it is simply the Wigner distribution of the transmission mask function $t(x)$: 
\begin{equation}
W(x,u)=\int t\left(x+\frac{x'}{2}\right)t^{*}\left(x-\frac{x'}{2}\right)e^{-2\pi ix'u}du
\label{eqn9}
\end{equation}
The Wigner distribution exhibit similar property as light fields: 1) $x$--shear transform for free--space propagation and 2) projection along the spatial frequency yields intensity. 
It is important to note that we are concerned with holograms under coherent light, and thus ray spaces of coherent fields. One important property of these ray spaces is their inclusion of negative values. Some of the flexibility of holograms to create complex 3D images can be thought of in terms of these negative values. The integration to intensity 
can change much more rapidly over short distances given access to negative values, unlike a parallax barrier with only positive rays.

\begin{figure}[tb]
  \centering
  \includegraphics[width=.95\linewidth]{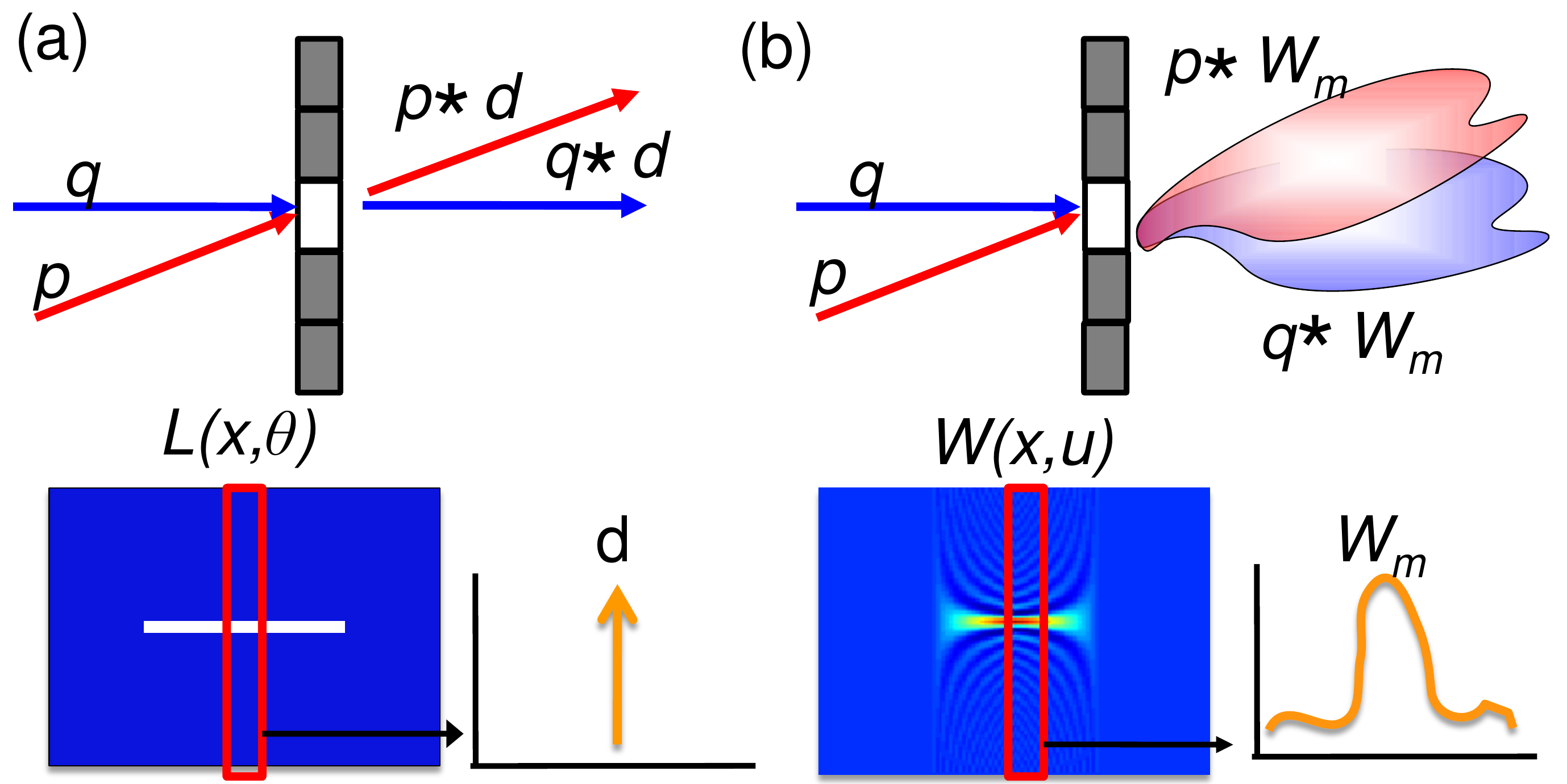}

  \caption{
           The operation of a thin mask under geometric optics considerations and under physical optics considerations. A ray passing through a slit is convolved with a $\delta$-function (vertical slice of a horizontal bar in phase space). The same slice through the Wigner distribution leads to a function of finite width $W_{m}$. Convolution with this function leads to a diffraction blur that is independent of incoming ray angle. }
\label{Fig11}
\end{figure}

A second important property of light at small scales is how it changes at the interface of a second mask, shown in Figure \ref{Fig11}. Since diffraction effects are now permitted, the angle of rays bend instead of just being attenuated. The amount of bending depends upon the ray space distribution of the mask itself, $W_{mask}(x,u)$. The final ray space distribution $W_{out}(x,u)$ is described through a convolution along the angular direction, 
\begin{equation}
W_{out}(x,u) = W_{in}(x,u)*|_{u} W_{mask}(x,u).
\label{eqn12}
\end{equation}
In the geometric limit of small wavelengths, the angular 'content' of a mask is negligible and can be expressed as a delta function. Returning to (\ref{eqn6}), we can return to the same multiplication relationship with,
\begin{equation}
l_{3}(x,\theta)=l_{1} * l_{2}(x,\delta_{\theta}) =l_{1}(x,\theta) l_{2}(x).
\end{equation}
This angular convolution is the fundamental difference in describing how rays change versus waves at a mask.


\subsection{Rank Comparison of Hologram and PB}

\begin{figure}[t!]
  \centering
  \includegraphics[width=.85\linewidth]{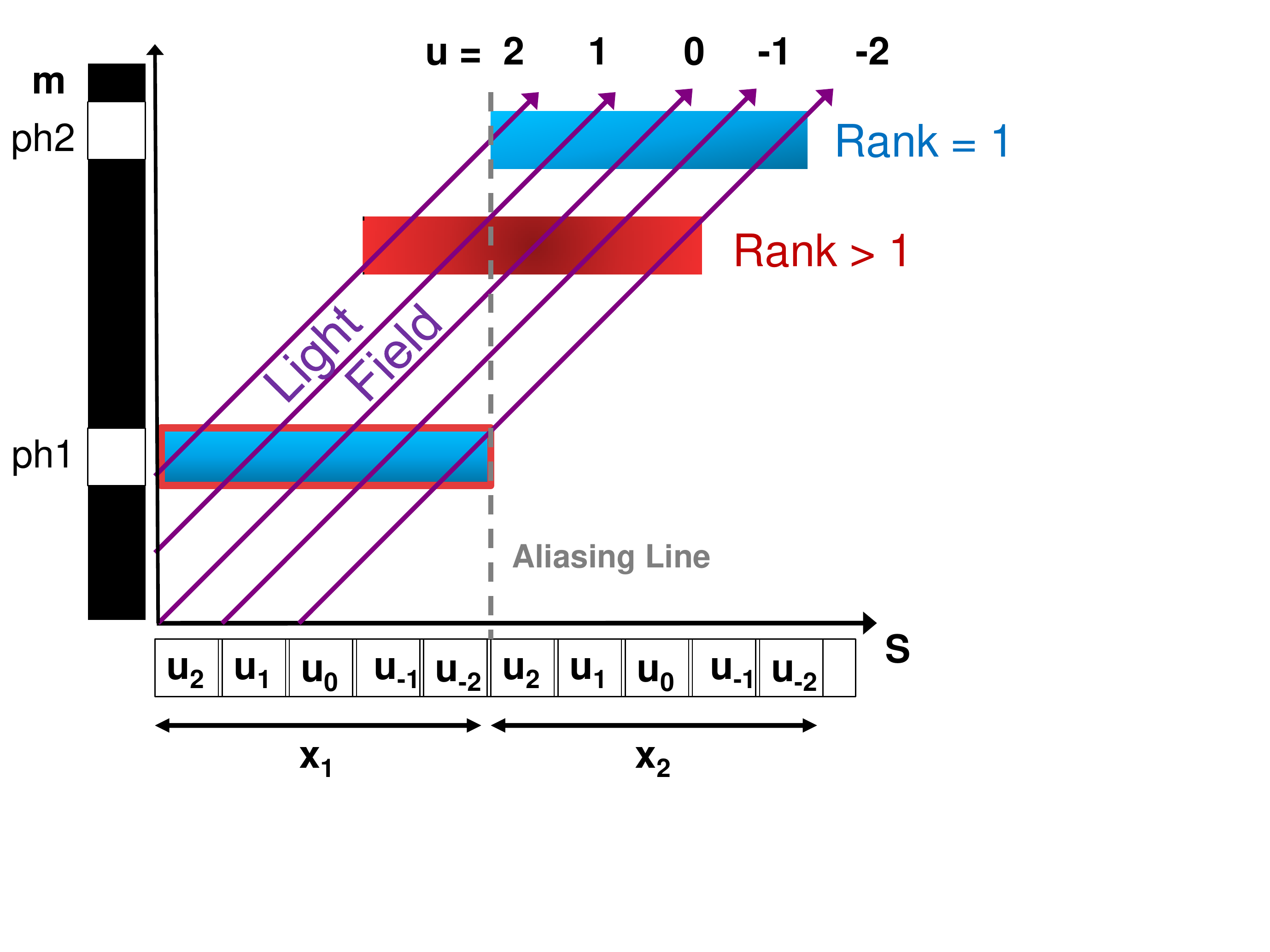}

  \caption{
        The light field from a parallax barrier (blue boxes) is decomposed into an outer product of screen coordinates $s$ and mask coordinates $m$, and is rank-1. The pinholes ($ph$) along the mask only allow certain strips of a continuous light field through.  The red box is an example of presenting an aliased light field, where a single pixel from screen $s$ is used to express two light field values.}
\label{Fig12}
\end{figure}

Similar decompositions in ray space will show the limits of each type of display. First considering a parallax barrier, we rewrite (\ref{eqn6}) in terms of the screen of a parallax barrier $s$ and the barrier mask $m$. The screen creates a broadband light field that propagates a distance $d$ to plane $m$. Using the shear relation
, we can write $l_{3}(x,\theta)$ in terms of a sheared $s$,
\begin{equation}
l_{3}(x,\theta)= s(x-d\theta)m(x) 
\label{eqn13}
\end{equation}
Thus, we see that the 2D light field from two amplitude planes is an outer product of 1D real column vectors:
 \begin{equation}
m(y_{1})s(y_{2})^{T}=[LF_{two planes}]
\end{equation}
This matrix has rank--1, which is quite a restrictive property. This decomposition is most clear when we present the light field in a rotated coordinate system, where the mask functions $m(y_{1})$ and $s(y_{2})$ are the axes (Figure \ref{Fig12}). Discrete lines of angular content $u$ at $45^{\circ}$ represent rays at different angles $\theta$ from $s$ through the direct relationship $u=d\theta$.
This implies that a parallax barrier undersamples any higher-rank light field it is designed to approximate. To avoid aliasing, it must sample the light field at large spatial intervals.  

The outer product in (\ref{eqn13}) is often how light field displays are generated using either a parallax barrier, lenticular array, or an integral imaging system. A specific consequence of parallax barrier displays is low light efficiency: to map a desired pixel to a location, all other rays blocked. This significant light attenuation may not necessarily be optimal. Other solutions to $m$ and $s$ could be achieved using linear algebra tools like single-value decomposition, which can force energy into few eigenvector components.

Turning to the ray space of a hologram, the easiest way to see how it is limited to a rank-1 condition is by examining the forward process of constructing the Wigner distribution. This begins with the mutual intensity function $J(x_{1},x_{2})=<t(x_{1}t^{*}(x_{2})>$, where the brackets denote an average. This function is a measure of the correlation between any two points $x_{1}$ and $x_{2}$ on a wavefront, or a mask $t$. Note that $J(x, x)$ is the intensity at $x$. For completely coherent illumination, any two points will be fully correlated. Thus, $J(x_{1},x_{2})$ can be expressed as an outer product ~\cite{Ozaktas:02} of the mask function $t(x_{1})$ with its complex conjugate $t^{*}(x_{2})$.

From this rank-1 space, we can create the familiar ray space picture first by rotating the coordinate system by $45^{\circ}$, and then re-scaling the $x$-axis a factor of 1/2. This is equivalent to a change of coordinates from $(x_{1},x_{2})$ to a new coordinate set $\left(x =\frac{x_{1}+x_{2}}{2}, x'=x_{1}-x_{2}\right)$. The Wigner distribution is just the Fourier transform of this rotated mutual intensity function along the angular axis. Its rank is increased significantly because only the Fourier transform along one axis ($x'$) is taken. However, the Wigner distribution remains a highly redundant representation. A coherent wavefront distribution after any propagation or modulation by multiple amplitude screens can always be created from an original rank-1 starting point. 
\begin{figure}[tb]
  \centering
  \includegraphics[width=.95\linewidth]{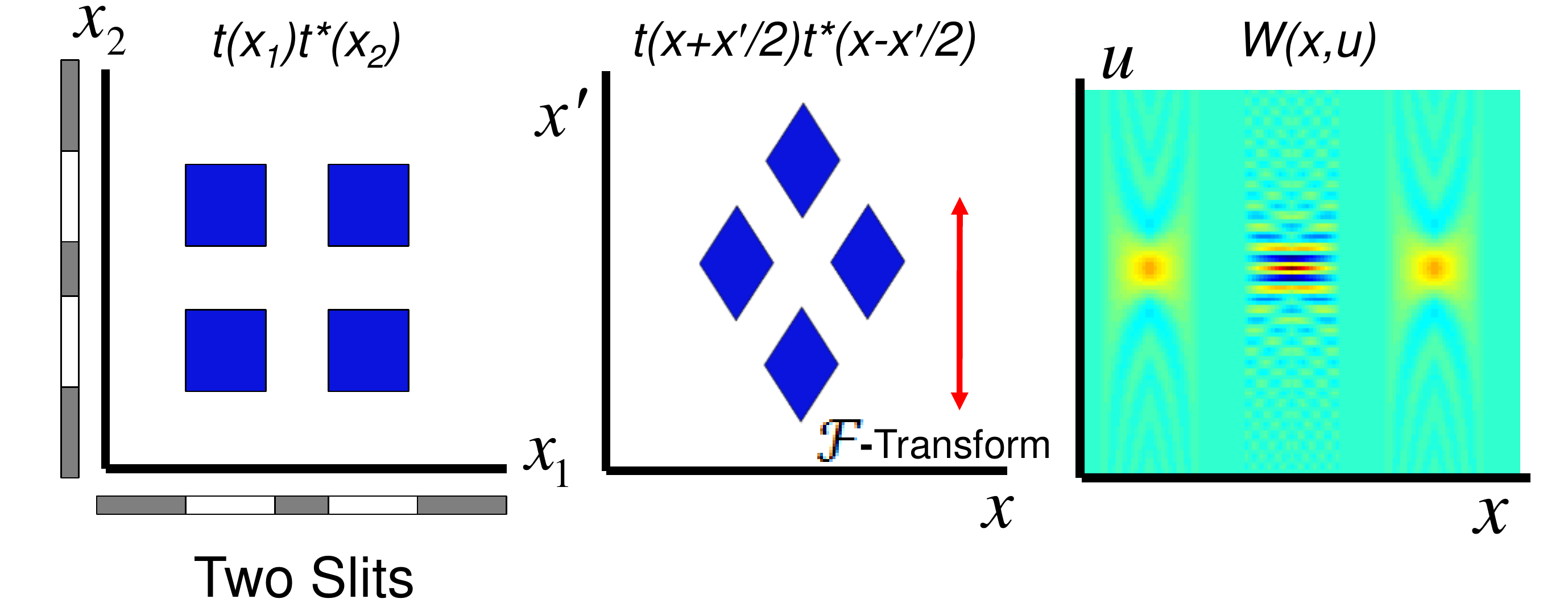}

  \caption{
           The ray space diagram for a hologram under coherent light assumptions can be created from a space that is originally rank-1. A rank-1 function in mutual intensity space transforms to a rotated coordinate system and then to the Wigner distribution.}
\label{Fig13}
\end{figure}

%


\subsection{The Larger Space of Light Fields}

Figure~\ref{Fig9} illustrates the entire hypothetical space of three--dimensional light distributions and methods for their generation. Given complete control over the light coherence state and the phase and amplitude of a mask, any arbitrary 3D distribution that obeys the wave propagation equation can be created. While this is an optimal situation for 3D display, there are a number of practical limitations. First is a physical limitation: the wave propagation equation is limited to smooth changes and out-of-focus noise can drastically decrease the SNR at a given viewing plane~\cite{Dorsch:94}. The second is a practical limitation: it is difficult to construct a light source with a given degree of partial coherence, and a thin sheet with both arbitrarily high grayscale and phase resolution. 

Using coherent light, there is a limited set of volumetric wavefronts that can be constructed. Any wavefront with a fixed coherence state must be definable over a single plane, since the distribution in all subsequent planes will be determined by the laws of propagation. Therefore, they must be able to be represented with a Wigner distribution, which we just found is rank-1. Within this space, there are even more limited sets of distributions that can be created with just amplitude, phase, or two planes of amplitude-only content. Examples of these limitations include an amplitude-only grating which always exhibits a zeroth order, or a phase-only plate which can not create a significant dark region at a short propagation distance. 

A two-plane display with incoherent light can create a set of distributions that overlaps with but differs from that of single plane under coherent light. For example, both can create two points as in Figure \ref{Fig10}, but only coherent light can create interesting patterns that require control over the phase of light, like an optical vortex. Having control over phase in a single plane becomes less important for display purposes as light becomes more incoherent. Therefore, the sizes and concentricity of spaces in Figure \ref{Fig9} change with coherence. 

Pixel size also scales with the axis of coherence. The most interesting feature of this analysis is the set of distributions that two amplitude modulation planes can create, and how it changes. The utility of a two-plane display like the parallax  barrier does not disappear when diffraction effects arise: it simply must be treated differently. As display feature sizes shrink, it seems that displays should begin to operate in a domain that moves away from completely incoherent light.

\section{Conclusion and Future Work}

We hope that some of connections made between the operation of parallax barriers and holograms stimulate further work at the intersection of the two display forms. With both a common framework to view resolution tradeoffs and also a shared rank--1 limitation in transform coordinates, there appears to be some significant overlap in functionality. At a minimum, we hope that approaching holograms from a discretized, more geometric based approach has offered some additional insight into their operation. 

Looking towards the future, it is possible to expand upon the rank analysis to design both incoherent, coherent, and even partially coherent-based displays that would be better suited for a specific display task. The inverse problem of moving from a particular desired light field to a mask and illumination source with a determined coherence state is the next step in this design process. Since the mutual intensity rank scales with coherence state for a single mask, light field distributions that otherwise could not be created with current displays may be generated considering partial coherence. Another direction would be to bring in lenticular displays into the analysis, where the angle of outcoming rays can be altered, instead of only considering their attenuation. 

In terms of future experimental work, we've taken part in preliminary discussion with museum curators interested in digitally preserving the largest collection of recorded holograms using our laser--scanning procedure. Furthermore, following the discrete Fourier hologram approach may allow for computationally simple and direct constructionof full Fresnel CGH holograms. Finally, purusing a hybrid form of display using both a holographic screen and parallax barrier layer may offer additional dimensions of design freedom.

%
%
%
%

\small
\bibliographystyle{acmsiggraph}
\bibliography{template}
\end{document}